\documentstyle[12pt,epsfig]{article}
\begin{document}
\bibliographystyle{stand}
\title{
Small-$x$ asymptotics of structure function ~$g_2$.
}
\author{
B.I.Ermolaev\thanks{Supported in part by the Grant
INTAS-RFBR-95-311} \\ {\em A.F.Ioffe Physico-Technical
Institute, St.Petersburg 194021, Russia} \\
\and
S.I.Troyan \\ {\em Petersburg Nuclear Physics Institute,
Gatchina 188350, Russia}
}

\date{}

\maketitle

\begin{abstract}
Nonsinglet structure function $g_2(x,Q^2)$ for deep
inelastic scattering of a lepton on a constituent quark is
calculated in the double logarithmic approximation at $x\ll~\!1$.
Small-$x$ asymptotics of $g_2$ is shown to have the same
singular behaviour as asymptotics of the nonsinglet structure
function $g_1$.  \end{abstract}

\section{Introduction}

Study of small-$x$ behaviour of structure functions
$g_1$ and $g_2$ of a constituent quark is essential for
understanding of polarization effects in deep
inelastic lepton-hadron scattering (DIS). Earlier $g_2$ has been studied
theoretically at $x\sim 1$ (see e.g.
\cite{BKL1,BKL2,BKLF,BB,ABH,BET,ALNR} and refs therein).

In framework of the conventional parton
model one has to consider the whole process of
DIS as a superposition of two
independent phases: hadron fragmentation in  constituent partons
and further deep inelastic scattering of the lepton on one of these
partons.

The first phase is controlled by non-perturbative (non-PT) QCD effects
and could be described today only in some model
approrach (see e.g. \cite{RR,DPPPW} and refs therein). The second phase
can be described in rigor terms of the perturbative (PT) QCD. The cross
section of the whole process then could be obtained as
convolution of the corresponding two probabilities.

At asymptotically high energies $s\to\infty$ behaviour of the cross
section is governed by the second phase, i.e. by the perturbative QCD,
while the first phase, related to hadron-scale distances, just
modifies pre-asymptotical factor and invokes strongly
suppressed higher twist corrections.

According to dispersion relations the DIS structure functions can be
presented as imaginary parts of the corresponding amplitudes for
forward compton scattering of the virtual photon on incoming hadron.

For unpolarized leptons and hadrons the cross section depends only
on crossing-symmetrical part of the compton amplitude. The
antisymmetrical part of the amplitude corresponds to
spin-dependent cross sections.

At asymptotically high energies the symmetrical part of the amplitude
strongly dominates over the antisymmetrical one. As a matter of fact
the leading log contributions come from Feynman diagrams mainly
constructed from gluon ladders. But for symmetrical case these
ladders are of BFKL-pomeron type~\cite{BFKL} whereas
for antisymmetrical case  they are simply of double log (DL)
type similar to the case of a non-singlet fermion ladder.  This
difference leads to an extra power factor of $1/x$ for
symmetrical amplitude.

The structure functions of unpolarized hadrons were extensively studied
both theoretically and experimentally by many authors. The
spin-dependent structure functions, though being more clear in
technique for theoretical analysis remain nonetheless less
studied.

The  spin-dependent part of the hadronic tensor,
$W^{[A]}_{\mu\nu}$, of the deep inelastic virtual photon --
quark scattering is defined by
\begin{equation}
W^{[A]}_{\mu \nu} = i\epsilon_{\mu\nu\alpha\beta} \frac{m
q_{\alpha}}{pq} \left( {\cal S}_{\beta}g_1 + \left({\cal S}_{\beta} -
p_{\beta}\frac{{\cal S}q}{pq} \right)g_2\right)
\label{WADEF}
\end{equation}

where $m$ and $\cal{S}$ are the mass and spin of the quark.
The functions $g_1$ and $g_2$ depend
on $x=Q^2/2pq$ and $Q^2$ $(Q^2= -q^2>0)$.
At $x\ll 1$ the Eq.(\ref{WADEF}) can be written in terms of
longitudinal and transversal projections of spin,
relative to the plane of the 4-vectors $p$ and $q$, in a
following way:
\begin{equation}
W^{[A]}_{\mu \nu} \simeq i\epsilon_{\mu\nu\alpha\beta} \frac{m
q_{\alpha}}{pq} \left( g_1{\cal S}^{\|}_{\beta} + (g_1
+ g_2){\cal S}^{\perp}_{\beta} \right)
\label{structures}
\end{equation}

 At very small-$x$, when $\ln(1/x) \gg \ln(Q^2/\mu^2)$, $\mu$ --
a mass scale which is order of few hundred $MeV$,  all
contributions $\sim (\alpha_s(\ln x)^2))^k$ are important and
must be taken into account. The structure functions
obtained  from DGLAP evolution equations \cite{GL,AP,D} did
not account for these terms as far as these equations were derived
originally for $x\sim 1$,  though analysis of the experimental data for
the polarised DIS within HERA energy range is perfectly compatible at
present \cite{ABFR} with extrapolating of DGLAP into the small $x$
region.

The double logarithmic asymptotics of $g_1$ for a constituent quark (a quark with
virtuality $\sim\mu^2$) was  calculated earlier \cite{BER1,BER2}
in the double-logarithmic approximation (DLA) for both
flavour non-singlet and singlet cases. Unlike $g_1$ the
perturbation series for $g_2$ begins  from higher
orders in QCD coupling $\alpha_s$. In particular, it was shown
\cite{ALNR} that there were no DL contributions to $g_2$ at small $x$
in the $\sim\alpha_s$ order.
We show below that in DLA the dominant contribution to $g_2$,
which corresponds to ladder graphs, can be obtained through
differentiation in QCD coupling  the analytical expression of
$g_1$.

This paper is organized as follows.

In Section 2 the pure quark ladder digarams are
considered and the analytical relation between
their contributions to $g_1$ and $g_2$ is derived.

In Section 3 the effects of non-ladder contributions
to $g_1$ and $g_2$ are discussed. Estimations for
the small-$x$ asymptotics of $g_1$ and $g_2$ are
presented.

\newpage
\section{Quark ladder  contributions}
\begin{figure}
\begin{minipage}[t]{5cm}
\centerline{\epsfig{file=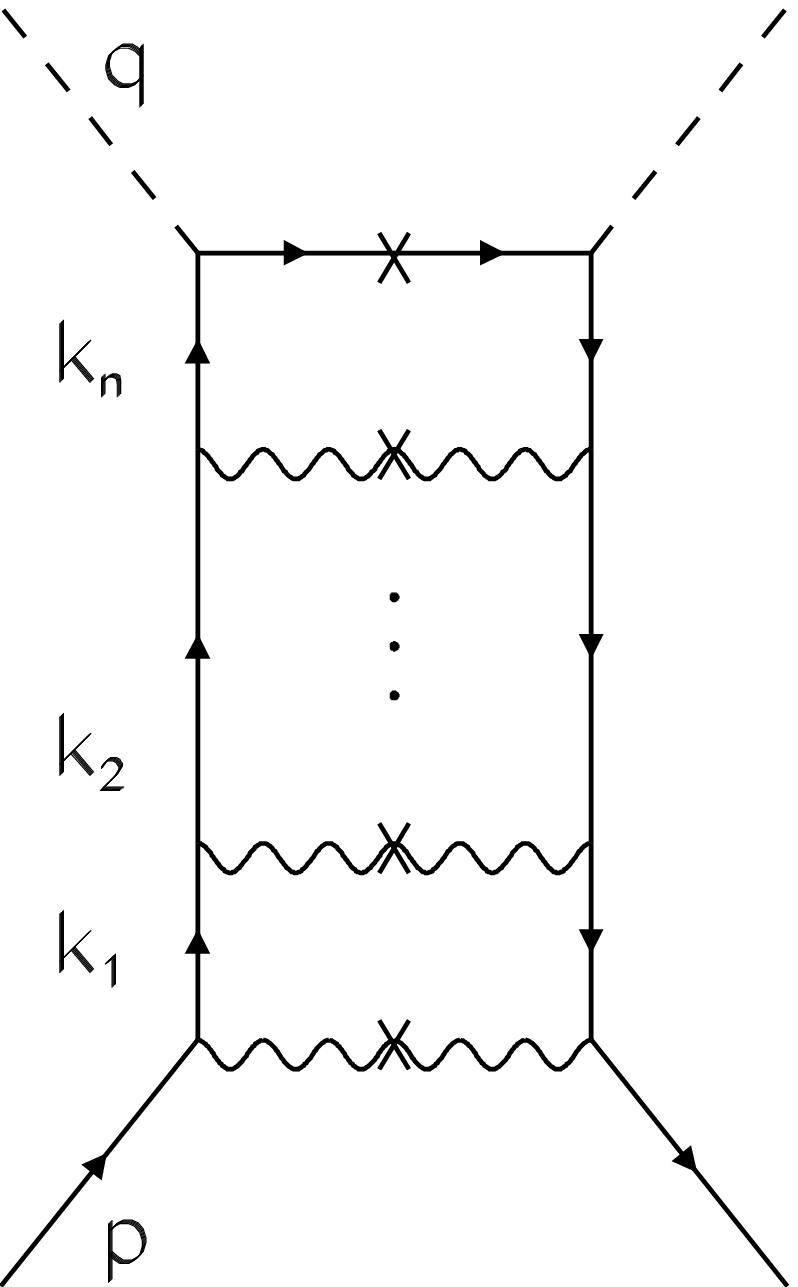,height=8cm}}
\caption{\label{qladder} A typical quark ladder graph.}
\end{minipage} \
\qquad \qquad \qquad \
\begin{minipage}[t]{5cm}
\centerline{\epsfig{file=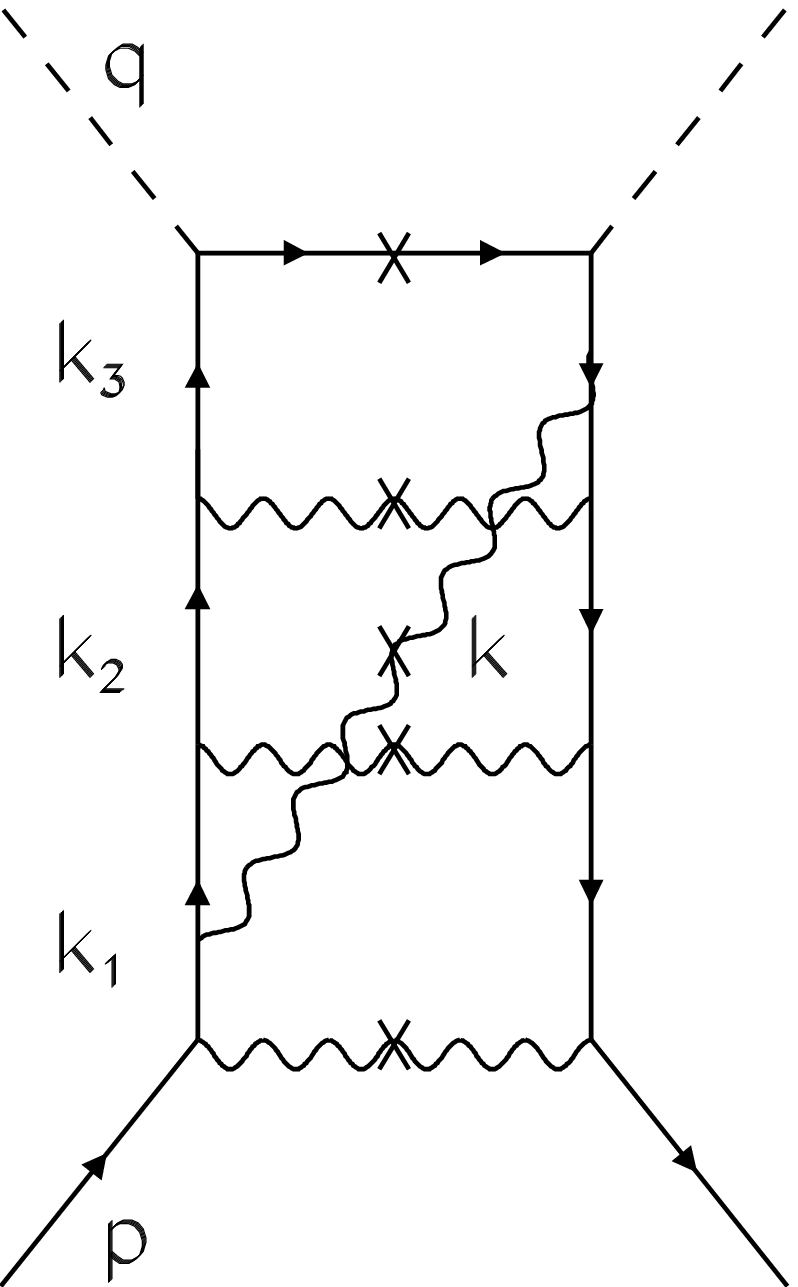,height=8cm}}
\caption{\label{nonladder} A typical nonladder graph.}
\end{minipage} \
\end{figure}

The quark ladder diagram with $n$ gluon rungs (see
Fig.(\ref{qladder})),

with a gluon propagator taken in the Feynman gauge
$$G_{\lambda\lambda^\prime}(k) = \frac{g_{\lambda\lambda^\prime}}{k^2 +
i\epsilon},$$
gives the following contribution to the general structure function
\begin{equation}\label{general}
W^{(n)}_{\mu\nu} = \frac{1}{2\pi}\!\mbox{Disc}\!
\int\!\!\ldots\!\int\prod^n_{i=1}\!\left[
\frac{d^4 k_i}{(2\pi)^4i}\frac{g_s^2C_F}{[k_i^2]^2}
\frac{- g_{\lambda\lambda^\prime}}{-
(k_i\!-\!k_{i\!-\!1})^2\!-\!i\epsilon} \right]\!
\frac{e^2\, T_{\mu\nu}^{(n)} }{-(q\!+\!k)^2\!-\!i\epsilon}
\end{equation}

\begin{equation}\label{trace}
T_{\mu\nu}^{(n)} = \frac{1}{2}\mbox{Tr}\left\{
(\hat{p}\!+\!m)(1\!-\!\gamma_5\hat{\cal S})
\gamma_{\lambda_1^\prime}\hat{k_1}\ldots\gamma_{\lambda_n^\prime}\hat{k_n}
\gamma_{\nu}(\hat{q}\!+\!\hat{k}_n)\gamma_{\mu}
\hat{k_n}\gamma_{\lambda_n}\ldots\hat{k_1}\gamma_{\lambda_1}
\right\}
\end{equation}

Here $g_s^2=4\pi\alpha_s$ is the QCD coupling constant,
$C_F=(N_c^2-1)/2N_c$ - the color factor and ``--'' signs are arranged
to make explicit the positive imaginary parts of propagators.  We have
omitted the quark mass $m$ in all internal quark propagators as only
linear in $m$ terms are essential and because of the well-known
relations
$$
\gamma_{\lambda}\gamma_{\mu}\gamma^{\lambda}= -2\gamma_{\mu} ,
\qquad
\gamma_{\lambda}\gamma_{\mu}\gamma_{\nu}
\gamma^{\lambda} = 4 g_{\mu\nu}
$$
and the restriction on a spin of an incoming on-shell quark,
$(p{\cal S})=0$, only the incoming quark $m$-term works.

To make the DL contribution explicit in Eq.(\ref{general}) let us
express all momentums in terms of the light-cone or Sudakov variables:
\begin{equation}\label{sudakov}
p=p^\prime + \frac{m^2}{s} q^\prime , \quad
q=-x p^\prime + q^\prime,  \quad s=2pq^\prime ,\quad
{p^\prime}^2={q^\prime}^2=0,
\end{equation}
$$
k_i=\beta_i p^\prime + \alpha_i q^\prime + k_i^{\perp} ,\quad
k_i^2=\beta_i\alpha_i s - k_{i\perp}^2,\quad
d^4k_i=d\beta_i d\alpha_i \frac{s}{2}d^2k_{i\perp}.
$$
Let us remind here that $x = Q^2/2pq^\prime$, $Q^2 = -q^2 > 0$.

We shall neglect the power suppressed terms ${\cal O}(m^2/s)$, ${\cal
O}(q^2/s)$ and thus ignore the difference between $p$, $q$ and
$p^\prime$, $q^\prime$ respectively in what follows. It is also
convenient to differ the Lorentz-covariant vector $k_i^{\perp}$ and
its two dimensional projection $\vec{k}_{i\perp}$: $(k_i^{\perp})^2 =
-\vec{k}_{i\perp}^2$.

Taking integrations over all $\alpha_i$ as residues in the
poles of gluon propagators,
\begin{equation}\label{alphas}
s\alpha_i = \frac{-
(\vec{k}_i\!-\!\vec{k}_{i\!-\!1})_{\perp}^2}{
\beta_{i\!-\!1}\!-\!\beta_i} + s\alpha_{i\!-\!1} ,
\end{equation}
we see that DL contribution may come only from the integration over
the region
\begin{equation}\label{region}
1 \gg \beta_1 \gg \beta_2 \ldots \gg \beta_n \geq x ,\qquad
\frac{\vec{k}_{1\perp}^2}{\beta_1} \ll \frac{\vec{k}_{2\perp}^2}{\beta_2}
\ldots \ll \frac{\vec{k}_{n\perp}^2}{\beta_n} ,
\end{equation}
inside which the whole ladder factorizes in the product of independent
rungs with quark virtualities
\begin{equation}\label{virtual}
k_i^2 = -\vec{k}_{i\perp}^2 -
\frac{\beta_i}{\beta_{i\!-\!1}}
\left(\vec{k}_{i\perp}\!- \vec{k}_{i\!-\!1\, \perp}\right)^2 \approx
-\vec{k}_{i\perp}^2 .
\end{equation}

Thus Eq.(\ref{general}) turns into
\begin{equation}\label{DLladd}
W_{\mu\nu}^{(n)}=\prod_{i=1}^n \left[\!\frac{g_s^2C_F}{16\pi^2}
\int_x^{\beta_{i-1}}\!\frac{d\beta_i}{\beta_i}
\int_0^{2\pi}\!\frac{d\phi_i}{2\pi}
\int_{\kappa_i}^{\beta_i s}\!\frac{dk_{i\perp}^2}{[k_i^2]^2}
\right] \beta_n\frac{e^2}{2s}
\delta(\beta_n\!-\! x\!-\!\frac{\vec{k}_{n\perp}^2}{s}) T_{\mu\nu}^{(n)} ,
\end{equation}
with
$$\kappa_i = \mu^2+\frac{\beta_i}{\beta_{i-1}}\vec{k}_{i-1 \perp}^2 $$
where the infrared cutoff $\mu^2$ was added to the lower boundary of
each transverse momentum to account for color bleaching which turns on
at confinement distances. The upper boundary for $\vec{k}_{i\perp}^2$ stems
from the requirement $(k_i+q)^2 \geq 0$.

To obtain the logarithmic integration over each $k_{i\perp}^2$ we
evidently have to see that $T_{\mu\nu}^{(n)}$ defined by the
Eq.(\ref{trace}) combined with the product of all denominators is
proportional to the product of all $k_{i\perp}^{-2}$ multiplied by some
dimensionless tensor which is independent of all $\beta_i$ and
$k_{i\perp}^2$.

In the region Eq.(\ref{region}) one can neglect the extra term
$\hat{k}_n$ which appears together with $\hat{q}$ in the middle of the
trace Eq.(\ref{trace}).  Then invoking a very useful and constructive
formula
\begin{equation}\label{usefull}
\gamma_{\nu}\hat{q}\gamma_{\mu} = i\varepsilon_{\mu\nu\lambda\sigma}
q^{\lambda} \gamma_5\gamma_{\sigma} + q_{\nu}\gamma_{\mu} +
q_{\mu}\gamma_{\nu} - g_{\mu\nu}\hat{q}
\end{equation}
immediately splits the tensor $T_{\mu\nu}^{(n)}$ into sum of symmetrical
and antisymmetrical parts:
\begin{eqnarray}\label{split}
T_{\mu\nu}^{(n)} &=& T_{ \{\mu\nu\} }^{(n)} + T_{[\mu\nu]}^{(n)} , \\
T_{ \{\mu\nu\} }^{(n)} &=& ( q_{\nu}g_{\mu\sigma} \!+\!
q_{\mu}g_{\nu\sigma} \!-\! q_{\sigma}g_{\mu\nu} ) \frac{1}{2}{\em
Tr}\left\{ (\hat{p}\!+\!m)(1\!-\!\gamma_5\hat{\cal S}) \Gamma^{\sigma}
\right\}, \nonumber\\ T_{[\mu\nu]}^{(n)} &=&
i\varepsilon_{\mu\nu\lambda\sigma} q^{\lambda} \frac{1}{2}{\em
Tr}\left\{ (\hat{p}\!+\!m)(1\!-\!\gamma_5\hat{\cal S})
\gamma_5\Gamma^{\sigma} \right\}, \nonumber
\end{eqnarray}
where
\begin{equation}\label{biggam}
\Gamma^{\sigma} = \left(\prod_{i=1}^n
[-g^{\lambda_i{\lambda_i}^{\prime}}] \right)
\gamma_{\lambda_1^\prime}\hat{k_1}\ldots\gamma_{\lambda_n^\prime}\hat{k_n} \gamma^{\sigma}
\hat{k_n}\gamma_{\lambda_n}\ldots\hat{k_1}\gamma_{\lambda_1}
\end{equation}
and the anticommutation property of $\gamma_5$ was used to take it out
of $\Gamma^{\sigma}$ for the antisymmetrical part.

The key point for understanding the structure of the matrix
$\Gamma_{\sigma}$ is that
\begin{equation}\label{ftensor}
\hat{k}\gamma^{\sigma}\hat{k} = - k^2 f^{\sigma\sigma^\prime}(k)
\gamma_{\sigma^\prime} , \qquad
f^{\sigma\sigma^\prime}(k) = g^{\sigma\sigma^\prime} -
2\, \frac{k^{\sigma} k^{\sigma^\prime}}{k^2} .
\end{equation}

This means that $\Gamma_{\sigma}$ is linear in $\gamma^{\sigma}$ and
can be expressed in terms of convolution of $n$ similar $f$-tensors:
\begin{equation}\label{azimut}
\Gamma^{\sigma} = \left(\prod_{i=1}^n \left[ - 2k_i^2 \right]\right)
{\cal E}^{\sigma}_{\tau}\gamma^{\tau} , \qquad
{\cal E}_{\sigma\tau} = \left({\prod_{i=1}^n}{\otimes} f(k_i)
\right)_{\sigma\tau}.
\end{equation}

According to Eq.(\ref{virtual}) the dimensional factor in
Eq.(\ref{azimut}) makes all integrations over $k_{i\perp}^2$ in
Eq.(\ref{DLladd}) the log integrals only if the integral over all
azimuthal angles
\begin{equation}\label{atens}
E^{(n)}_{\sigma\tau} =
\int_{0}^{2\pi}\!\frac{d\phi_1}{2\pi}\!
\left[\frac{\vec{k}_{1\perp}^2}{-k_1^2}\right]
\ldots
\int_{0}^{2\pi}\!\frac{d\phi_n}{2\pi}\!
\left[\frac{\vec{k}_{i\perp}^2}{-k_i^2}\right]
 \,{\cal E}^{(n)}_{\sigma\tau}
\end{equation}
does not implicitly depend on $k_{i\perp}^2$ or $\beta_i$.
We shall prove this later.
Substituting Eq.(\ref{atens}) and Eq.(\ref{split}) into
Eq.(\ref{general}) we finally obtain the following compact formulae
which describe all structure functions corresponding to the considered
ladder diagram:
\begin{eqnarray}\label{final}
W_{\{\mu\nu\}}^{(n)} &=& - g_{\mu\nu}^{\perp}M^{(n)}
\left(E^{(n)}\right)_{\sigma\rho} \, \frac{q^{\sigma}p^{\rho}}{(pq)} \\
W_{[\mu\nu]}^{(n)} &=& i\varepsilon_{\mu\nu\lambda\sigma}
\frac{mq^{\lambda}}{(pq)} M^{(n)} \left(E^{(n)}\right)^{\sigma}_{\rho}
{\cal S}^{\rho}
\end{eqnarray}
where
\begin{equation}\label{moper}
M^{(n)} = \left(\frac{\alpha_s C_F}{2\pi}\right)^n \prod_{i=1}^n
\left[ \int_{x}^{\beta_{i-1}}\frac{d\beta_i}{\beta_i}
\int_{\kappa_i}^{\beta_i s}\frac{dk_{i\perp}^2}{k_{i\perp}^2} \right]
\beta_n \frac12 e^2 \delta(\beta_n\!-\!x\!-\!\frac{k_{n\perp}^2}{s})
\end{equation}
can be considered as a DL integral operator.

Due to the trivial relation
\begin{equation}\label{averone}
\int_0^{2\pi}\!\frac{d\phi}{2\pi} k_{\sigma}^{\perp} k_{\rho}^{\perp}
= (k^{\perp})^2 g_{\sigma\rho}^{\perp}
\end{equation}
the integral
\begin{equation}\label{key}
\int_{0}^{2\pi} \frac{d\phi}{2\pi} f_{\sigma\sigma^\prime}(k^{\perp}) =
g_{\sigma\sigma^\prime}^{\|}
\end{equation}
in the DL region acts like a projector on the longitudinal subspace.
If one might neglect the longitudinal components of the momenta $k_i$,
$k_i^{\|} = \beta_i p + \alpha_i q$, then we would immediately obtain
the obvious results:
\begin{equation}\label{trivial}
E^{(n)}_{\sigma\rho} p^{\rho} = p_{\sigma}, \qquad
E^{(n)}_{\sigma\rho} {\cal S}_{\|}^{\rho} = {\cal S}^{\|}_{\sigma},
\qquad
E^{(n)}_{\sigma\rho} {\cal S}_{\perp}^{\rho} = 0.
\end{equation}

In reality the account of the longitudinal momenta components does
not influence the first two results of Eq.(\ref{trivial}) but changes
the last prediction dramatically.  To see this let us present ${\cal
E}$ of Eq.(\ref{azimut}) as follows:
\begin{equation}\label{decompos}
{\cal E}_{\sigma\rho}^{(n)} = g_{\sigma\rho} - \sum_{i=1}^n
2\frac{(k_i)_{\sigma}^{\perp} (k_i)_{\rho}}{k_i^2} + \sum_{i>j}^n
2\frac{(k_i)_{\sigma}^{\perp} 2(k_i k_j)(k_j)_{\rho}}{k_i^2 k_j^2} + \ldots
\end{equation}

We have made it clear here that only $(k_i^{\perp})_{\sigma}$ component
can work in Eq.(\ref{final}), as the $q$-component gives
zero or power suppressed contribution and the $p$-component leads
beyond the DL approximation because of the small factor $\beta_i$.

The first two terms of Eq.(\ref{decompos}) obviously contribute to
Eq.(\ref{final}) as
\begin{equation}\label{eresult}
E_{\sigma\rho}^{(n)} = g_{\sigma\rho} - n g_{\sigma\rho}^{\perp}
\end{equation}

One might expect to get some contribution from non-diagonal terms
$(k_i)_{\sigma}^{\perp}\alpha_i q_{\rho}$ as all $\alpha_i$ depend on
azimuthal angles according to Eq.(\ref{alphas}). But
\begin{equation}\label{avertwo}
\int_0^{2\pi}\!\frac{d\phi_i}{2\pi} (k_i^{\perp})_{\sigma} \alpha_i
q_{\rho} = -(k_{i\!-\!1}^{\perp})_{\sigma}
\frac{(k_i^{\perp})^2}{\beta_{i\!-\!1} s} q_{\rho}
\end{equation}
and subsequent integration over $\phi_{i\!-\!1}$ turns this contribution
to zero. To be correct at  such circumstances one must do not forget to
include the small azimuthal dependence of $k_i^2$ (see Eq.(\ref{virtual}))
in denominators in Eq.(\ref{atens}) into consideration:
\begin{equation}\label{averthree}
\int_0^{2\pi}\!\frac{d\phi_{i\!-\!1}}{2\pi}
\frac{(k_{i\!-\!1}^{\perp})_{\sigma}}{k_{i\!-\!1}^2} =
\frac{\beta_{i\!-\!1}}{\beta_{i\!-\!2}}
\frac{(k_{i\!-\!2}^{\perp})_{\sigma}}{(k_{i\!-\!1}^{\perp})^2}
\end{equation}
Performing the chain of subsequent integrations over azimuth angles we
come eventually to the last integral over $\phi_1$ which turns to zero.

The third term of Eq.(\ref{decompos}) despite its large magnitude
$$
-2(k_i k_j)=-2(k_i^{\perp}k_j^{\perp})-\beta_j\alpha_i s \approx
-\frac{\beta_j}{\beta_{i\!-\!1}}(k_i^{\perp}\!-\!
k_{i\!-\!1}^{\perp})^2 \gg -(k_i^{\perp})^2 .
$$
does not contribute in Eq.(\ref{final}) at all. As a matter of fact,
according to Eqs.(\ref{averone}-\ref{averthree})
\begin{equation}\label{averfour}
\int_0^{2\pi}\!\frac{d\phi_i}{2\pi_i} (k_i^{\perp})_{\sigma} 2(k_i k_j)
(k_j)_{\rho} = (k_i^{\perp})^2 \left(
(k_j^{\perp})_{\sigma} -
\frac{\beta_j}{\beta_{i\!-\!1}} (k_{i\!-\!1}^{\perp})_{\sigma} \right)
\end{equation}
and subsequent integrations over $\phi_{i\!-\!1}$, $\phi_{i\!-\!2}$, \ldots
$\phi_{j\!+\!1}$ transform the result of Eq.(\ref{averfour}) into
$$
(k_j^{\perp})_{\sigma} -
\frac{\beta_j}{\beta_{i\!-\!1}}
\frac{\beta_{i\!-\!1}}{\beta_{i\!-\!2}} \cdots
\frac{\beta_{j\!+\!1}}{\beta_j}
(k_j^{\perp})_{\sigma} = 0 .
$$

As soon as large tensor magnitude was integrated down to zero the
problem of dangerous small corrections to it arises again. We omit the
cumbersome analysis of such corrections here and just present the
result: the third term of Eq.(\ref{decompos}) indeed contribute nothing
in DL approximation.

With this result the next term which is not shown in Eq.(\ref{decompos})
and can be expressed as the convolution of already considered terms
gives also zero contribution. Therefore Eq.(\ref{eresult}) can be considered
as the total expression for substitution into Eq.(\ref{final}):
\begin{eqnarray}\label{result}
W_{\{\mu\nu\}}^{(n)} &=& - g_{\mu\nu}^{\perp}M^{(n)} \nonumber \\
W_{[\mu\nu]}^{(n)} &=& i\varepsilon_{\mu\nu\lambda\sigma}
\frac{mq^{\lambda}}{(pq)} \left\{ M^{(n)}{\cal S}^{\sigma} -
\frac{\partial}{\partial\ln\alpha_s} M^{(n)} {\cal
S}_{\perp}^{\sigma} \right\} \end{eqnarray}
where we have used the simple relation
$$
\frac{\partial\alpha_s^n}{\partial\ln\alpha_s}  = n \alpha_s^n
$$
to express the transverse spin contribution in terms of the amplitude
$M^{(n)}$ of Eq.(\ref{moper}).

Summing up over all ladders thus leads to a very simple relation
between the contributions to DIS functions $g_1$ and $g_2$:
\begin{equation}\label{ladreltn}
g_2 = - \frac{\partial g_1}{\partial\ln\alpha_s}
\end{equation}

\section{Discussion}
Eq.(\ref{ladreltn}) is valid  only for ladder Feynman graphs.
Though these diagrams determine the small-$x$ asymptotics of the DGLAP
evolution equations, this is not true for $g_1$ and $g_2$ in the DL
region (\ref{region}).

To understand the effect of nonladder
graphs on Eq.(\ref{ladreltn}), let us add a nonladder gluon to a
ladder graph (see Fig.(\ref{nonladder})). Nonladder gluon momentum $k$
must be soft enough not to destroy the DL pattern of a ladder graph.
Unlike a skeleton ladder gluon a nonladder gluon invokes an additional
DL integration only if no any $k_{\perp}^2$ factor appears in the
integrand's numerator. Such factor would cancel the softest  virtuality
propagator (of those invoked by a nonladder gluon) in the denominator.
Thus adding a nonladder gluon does not change the relation between
 contributions of the ${\cal S}^{\perp}$ and ${\cal
S}^{\|}$-structures, i.e. the Eq.(\ref{eresult}).

Therefore we could use the Eq.(\ref{ladreltn}) if one would separate
out the contribution of ladder skeleton gluons to $g_1$ and
differentiate it in $\alpha_s$. Fortunately one can use the
expression for $g_1$ obtained in the work~\cite{BER1} and tag the
ladder and nonladder gluon contributions through definition of separate
QCD couplings $\alpha_{L}$ and $\alpha_{NL}$ respectively, $\alpha_{L}
= \alpha_{NL} = \alpha_s$:
\begin{eqnarray} \label{g1}
g_1 &=& \frac{e^2_q}{2}\,\int_{-i\infty}^{i\infty}\!\frac{d\omega}{2\pi
i} \left( \frac{1}{x} \right)^{\omega} \frac{\omega}{\omega \!-\!
f_0} \left( \frac{Q^2}{\mu^2} \right)^{f_0}  \\ f_0 &=& \frac12 \left[
\omega \!-\! \sqrt{\omega^2 \!-\! \frac{2\alpha_L C_F}{\pi} \!+\!
\frac{4\alpha_{NL}^2 C_F N_c}{\omega\pi^2} V} \right] \nonumber \\ V
&=& \frac{d}{d\omega}\ln \int_0^{\infty} \! dt\ t^{-(1\!+\!p)}
\exp{\left(-\frac{t^2}{2} \!-\!
\frac{t\omega}{\sqrt{\frac{\alpha_{NL}N_c}{2\pi}}}\right)}.  \nonumber
\end{eqnarray}
where $V$ defines the nonladder contribution, $p=-1/2N_c^2$.

Now Eq.(\ref{ladreltn}) can be written in the form
\begin{equation} \label{finalres}
g_2 = -\left.\frac{\partial g_1}{\partial\ln\alpha_L}\right|_{\alpha_L =
\alpha_{NL} = \alpha_s} .
\end{equation}
Eqs.(\ref{g1},\ref{finalres}) show  the same asymptotical
behaviour of $g_1$ and $g_2$  at $x\to 0$:
\begin{equation} \label{as}
g_2 \sim g_1\sim \left( \frac{1}{x}\right)^a
\left(\frac{Q^2}{\mu^2} \right)^{\frac{a}{2}}, \qquad
a\simeq\sqrt{\frac{2\alpha_s C_F}{\pi}}
\left( 1\!+\!\frac{1}{2N_c^2} \right) .
\end{equation}

\section{Conclusion}
We have considered only nonsinglet contribution to the structure
function $g_2$. Obviously, one should expect the singlet contribution,
i.e. insertion of gluon ladders, to be dominating over the nonsinglet
one -- similar to the case of the structure function $g_1$~\cite{BER2}.
Our preliminary consideration has shown that one could repeat the
analysis presented above for the case of gluon ladder contributions,
the crucial points of the analysis as well as general features of the
result remain the same. The only difference stems from different
contributions of a gluon rung to the numerator's tensor structure
(\ref{atens}) for each case: for a quark ladder - Eq.(\ref{ftensor})
and for a gluon ladder -
$$
h^{\sigma\sigma^{\prime}}(k) = g^{\sigma\sigma^{\prime}} -
\frac{k^{\sigma}k^{\sigma^{\prime}}}{k^2}.
$$

Thus one would expect the following generalization of the expression
(\ref{finalres}) for the total structure function $g_2$:
$$
g_2 = -\left.\frac{\partial
g_1}{\partial\ln\alpha_L}\right|_{\alpha_L = \alpha_{NL} =
\alpha_s} -\ \frac12\left.\frac{\partial
g_1}{\partial\ln\widetilde{\alpha}_L}\right|_{\widetilde{\alpha}_L =
\alpha_{L} = \alpha_{NL} = \alpha_s}.
$$
where we discern the QCD coupling $\widetilde{\alpha}_L$ corresponding
to gluon radiation in a gluon cell from the QCD coupling
$\alpha_L$
corresponding to gluon radiation in a quark cell of a ladder graph.

A detailed comprehensive consideration which is necessary to prove this
relation will be published elsewhere.

\vspace{0.5cm}
We are thankful to A.P.Bukhvostov for the fruitful discussion.

\end{document}